\documentclass[useAMS,usenatbib]{mn2e}

\usepackage{graphicx}
\usepackage{rotating}

\title[On the diffuse X-ray emission from the WR bubble NGC\,2359]{On
  the diffuse X-ray emission from the Wolf-Rayet bubble NGC\,2359}
\author[Toal\'{a} et al.]{J.A.\,Toal\'{a}\thanks{E-mail:toala@iaa.es}$^{1}$, M.A.,\,Guerrero$^{1}$, Y.-H.\,Chu$^{2}$, and R.A.\,Gruendl$^{2}$\\
$^{1}$Instituto de Astrof\'{i}sica de Andaluc\'{i}a, IAA-CSIC, Glorieta de la 
Astronom\'{i}a s/n, 18008 Granada, Spain\\
$^{2}$Department of Astronomy, University of Illinois, 1002 West Green Street, Urbana, IL 61801, USA
}

\def\gtsima{$\; \buildrel > \over \sim \;$}    
\def\gtrsim{\lower.5ex\hbox{\gtsima}}           
\def\lesssim{\lower.5ex\hbox{\ltsima}}           
\def\ltsima{$\; \buildrel < \over \sim \;$}    

\begin{document}


\maketitle

\label{firstpage}

\begin{abstract}

  A recent {\it XMM-Newton} observation (Zhekov\,2014) has revealed
  diffuse X-ray emission inside the nebula NGC\,2359 around the
  Wolf-Rayet star WR\,7. Taking advantage of an improved point-source
  rejection and background subtraction, and a detailed comparison of
  optical and X-ray morphology, we have reanalyzed these X-ray
  observations. Our analysis reveals diffuse X-ray emission from a
  blowout and the presence of emission at energies from 1.0 to
  2.0~keV. The X-ray emission from NGC\,2359 can be described by an
  optically-thin plasma emission model, but contrary to previous
  analysis, we find that the chemical abundances of this plasma are
  similar to those of the optical nebula, with no magnesium
  enhancement, and that two components at temperatures
  $T_{1}$=2$\times$10$^{6}$~K and $T_{2}$=5.7$\times$10$^{7}$~K are
  required. The estimated X-ray luminosity in the 0.3--2.0~keV energy
  range is $L_\mathrm{X}$=2$\times$10$^{33}$~erg~s$^{-1}$. The
  averaged rms electron density of the X-ray-emitting gas
  ($n_\mathrm{e}\lesssim$0.6~cm$^{-3}$) reinforces the idea of mixing
  of material from the outer nebula into the hot bubble.

\end{abstract}


\section{Introduction} 
\label{sec:intro}

High-quality X-ray observations ({\it Chandra}, {\it XMM-Newton}, and
{\it Suzaku}) have been performed towards five WR nebulae, namely
S\,308, NGC\,2359, RCW\,58, and NGC\,6888 (around WR\,6, WR\,7,
WR\,40, and WR\,136, respectively) and that around WR\,16, but diffuse
X-ray emission has been detected only in three of them \citep[S\,308,
NGC\,2359, and
NGC\,6888;][]{Chu2003,Gosset2005,Zhekov2011,Toala2012,Toala2013,Toala2014,Zhekov2014}.
In the cases of S\,308 and NGC\,6888, their X-ray spectra have been
modeled with two-temperature plasma emission. The first temperature
component is around 1-1.4$\times$10$^{6}$~K and the second
$\sim$10$^{7}$~K, which contributes $<$10\% of the total flux. The
diffuse X-ray emission in these two WR bubbles has been successfully
modeled using the optical nebular abundances, strongly indicative of
injection of material from the outer ionized nebula into the hot
interiors via instabilities created in the wind-wind interaction zone
and/or via thermal conduction
\citep[e.g.,][]{Arthur2007,Pittard2007,Toala2011,Dw2013}. The spectral
similarities do not translate to the spatial properties of the diffuse
emission: S\,308 displays a limb-brightened morphology, while
NGC\,6888 exhibits three maxima detected with {\it Chandra} and {\it
  XMM-Newton} \citep[][Toal\'{a} et al. in
prep.]{Toala2012,Toala2014}. It is, thus, imperative to increase the
number of detections of WR bubbles that harbor diffuse hot gas in
order to study the similarities (or differences), to study their
dependence on the structure of the ISM and physical parameters of
stellar winds that produce the hot plasma.


\begin{figure}
\begin{center}
\includegraphics[width=1.0\linewidth]{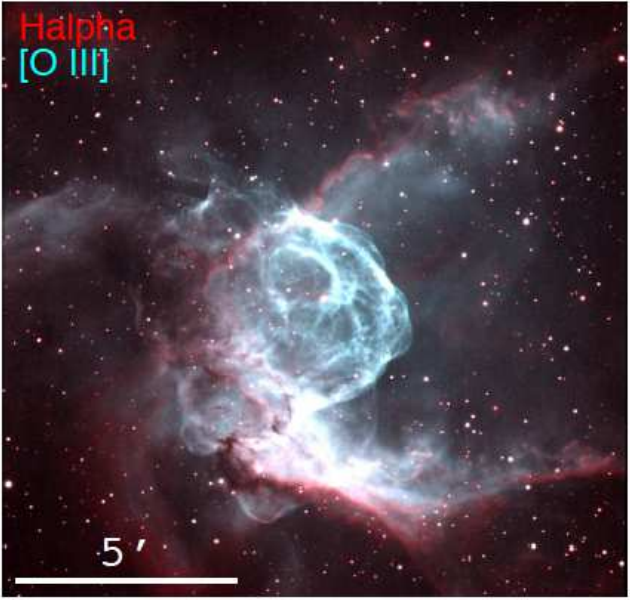}~
\end{center}
\caption{Color-composite optical image of NGC\,2359. Red and blue
  correspond to H$\alpha$ and [O\,{\sc iii}] line emission. Images
  were taken by the Stars Shadows Remote Observatory (SSRO) South and
  located at Cerro Tololo Inter-American Observatory. North is up,
  east to the left.}
\label{fig:image_ngc2359}
\end{figure}

The WR nebula NGC\,2359 around WR\,7 presents a filamentary central
bubble, blisters, and several filaments as seen in the optical image
presented in Figure~\ref{fig:image_ngc2359}. These characteristics
indicate that the central star has experienced a phase of strong and
episodic mass loss as a yellow supergiant and/or luminous blue
variable (LBV). Recently, \citet[][hereafter Z14]{Zhekov2014}
presented the discovery of diffuse X-ray emission towards this WR
bubble. The {\it XMM-Newton} observations were analyzed using the
XMM-ESAS tasks which are optimized for extended sources and apply very
restrictive criteria for the selection of events that reduce
considerably the net count number. The resulting spectrum was modeled
with a single plasma temperature ($kT$=0.21~keV;
$T$=2.4$\times$10$^{6}$~K) for abundances of the optical nebula with
anomalously enhanced magnesium abundances. 
The value of the reduced $\chi^{2}$ of the fit is smaller than unity, 
$\chi^{2}$/DoF=0.72, casting doubts on the validity of the fit as it 
suggests that the errors in the data have been over-estimated.  

Another problem that plagued the spectral fit presented by Z14 is 
the presence of an emission line at $\sim$1.14~keV, which has no 
identification and whose presence/absence depends on the selection 
of the background.

In this work we present a reanalysis of the {\it XMM-Newton}
observations of NGC\,2359. We show that point-like sources not removed
by Z14 and a questionable selection of the background hampered the
analysis of the diffuse X-ray emission. An additional patch of diffuse
X-ray emission towards the northeast from the central WR star was also
missed by that study. Our spectrum allows better spectral analysis up
to energies of 3.0~keV improving the quality of the spectral and
spatial characterizations and reliability of the analyses.

This paper is presented as follows. In Section~2 we give a short
summary of the {\it XMM-Newton} observations. Section~3 and 4 present
the distribution of the diffuse X-ray emission and the spectral
analysis, respectively. We discuss our findings in Section~5, and
finally conclude in Section~6.

\section{\textit{XMM-Newton} Observations}

The {\it XMM-Newton} observations of the WR nebula NGC\,2359 were
performed on 2013 April 9 (Obs.Id: 0690390101; PI: S.\,Zhekov) during
revolution 2442. The EPIC cameras (MOS1, MOS2, and pn) were operated
in the full-frame mode with the medium optical filter for total
exposure times of 110.8, 110.9, and 109.3~ks, respectively. 
The observations were processed using the \emph{XMM-Newton} Science
Analysis Software (SAS Ver 13.5.0) with the associated calibration
Files (CCF) available on 2014 June 16.  

\begin{figure*}
\begin{center}
\includegraphics[width=0.5\linewidth]{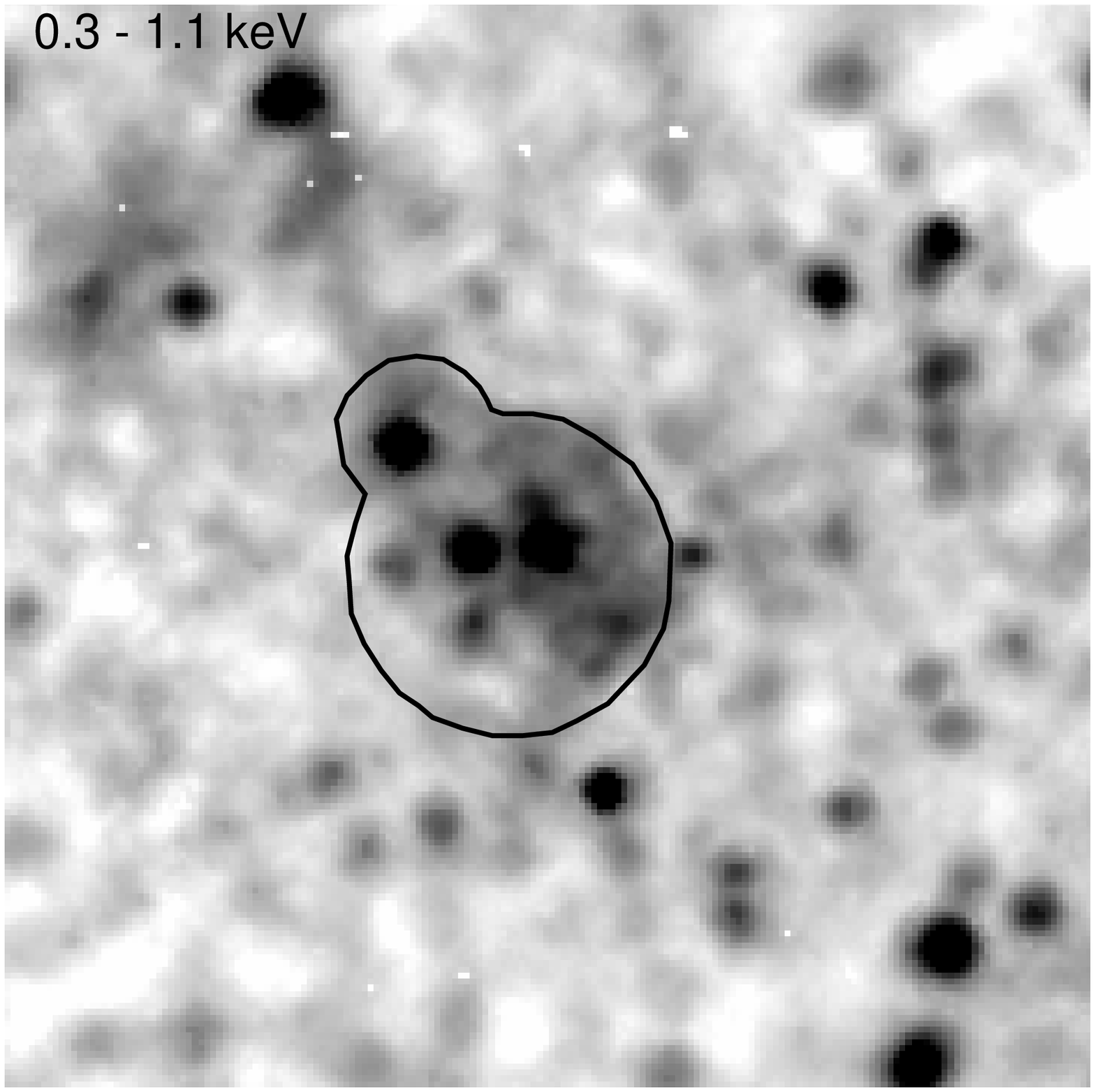}~
\includegraphics[width=0.5\linewidth]{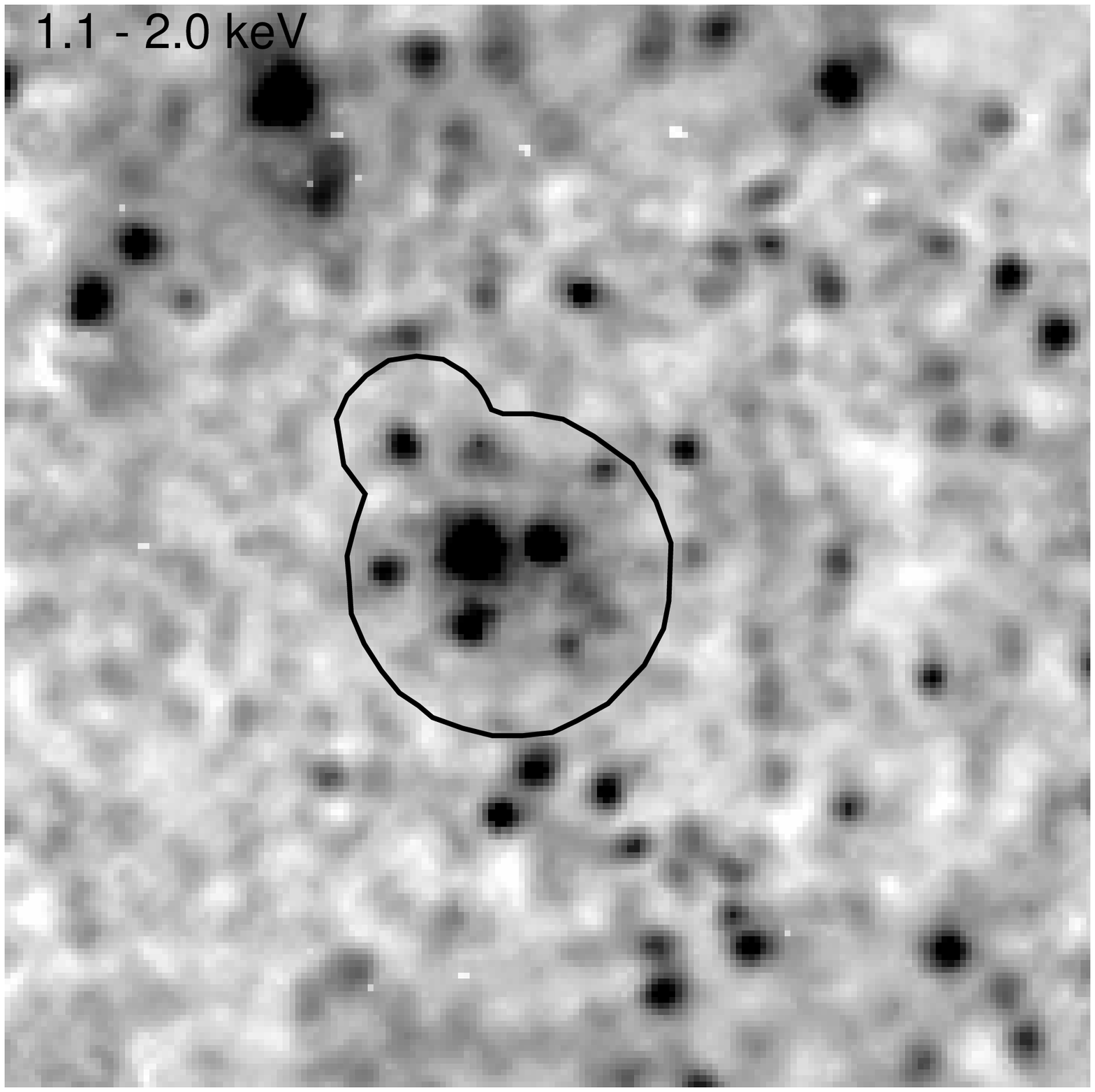}\\
\includegraphics[width=0.5\linewidth]{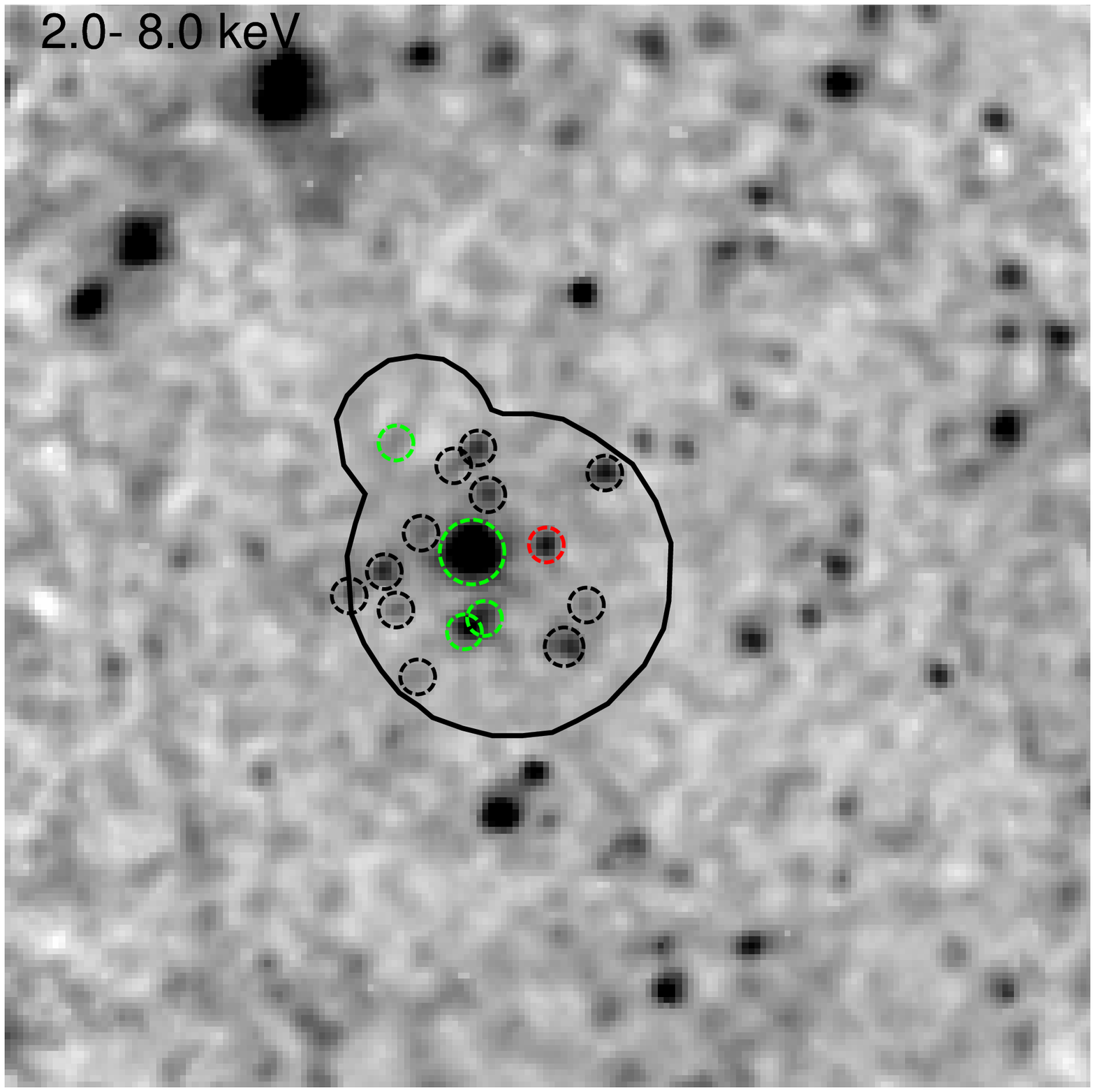}~
\includegraphics[width=0.5\linewidth]{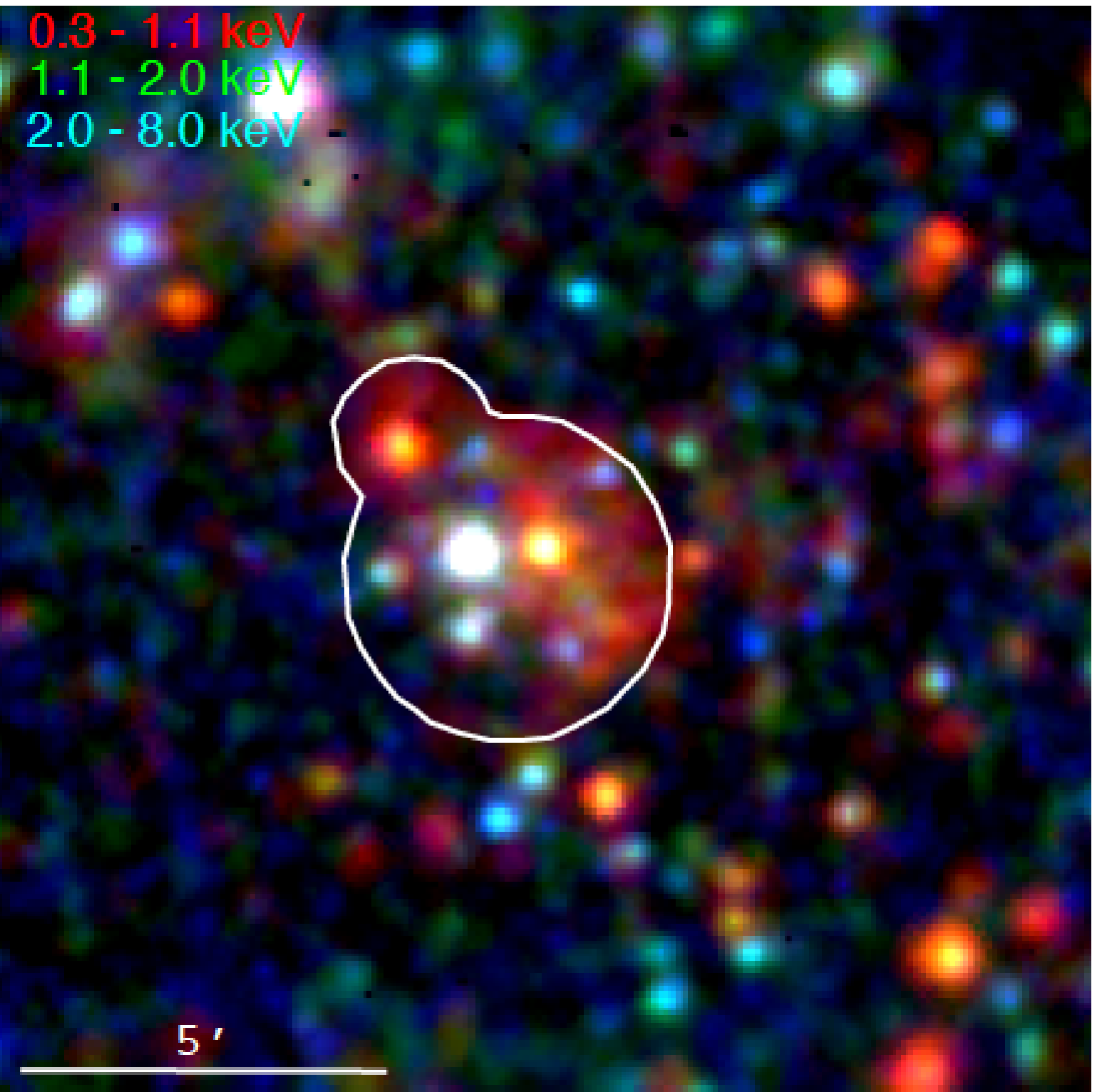}
\end{center}
\caption{{\it XMM-Newton} EPIC exposure-corrected,
  background-subtracted X-ray images in three different energy bands
  in the field of view of NGC\,2359. The energy bands are labeled in
  the upper left corner of each panel. The bottom-right panel shows a
  color-composite image of the three other panels. The diffuse X-ray
  emission towards NGC\,2359 is delimited by solid-line (black or
  white) polygonal aperture and the identified point-like sources with
  dashed-line circles (only shown in the bottom-left panel). The point
  sources excised by Z14 are shown in red (e.g., WR\,7) and
  green. North is up, east to the left.}
\label{fig:ngc2359_xrays}
\end{figure*}

We first analyzed the observations making use of the XMM-ESAS tasks to
produce images of NGC\,2359 in different energy bands (see
Section~\ref{sec:distribution}). This allowed us to identify the
distribution of the diffuse X-ray emission and the locations of
point-like sources.  We then used the SAS tasks \emph{evselect},
\emph{arfgen}, and \emph{rmfgen} to extract the X-ray spectrum of the
diffuse X-ray emission from the WR nebula and produce the associated
calibration matrices as described in the corresponding SAS threads
(Section~\ref{sec:spectral_a}).  This is justified because the diffuse
X-ray emission in NGC\,2359 does not fill the field of view of the
EPIC cameras.  Whereas the XMM-ESAS tasks apply very restrictive event
selection criteria which are appropriate for the analysis of the
spatial distribution of the X-ray emission, these are not required for
spectral analysis.



\section{Distribution of the X-ray emission}
\label{sec:distribution}

The X-ray images were produced following Snowden \& Kuntz's cookbook
for analysis of {\it XMM-Newton} EPIC observations of extended objects
and the diffuse background (Version 5.9) taking into account their
corresponding Current Calibration
Files\footnote{ftp://legacy.gsfc.nasa.gov/xmm/software/xmm-esas/xmm-esas-v13.pdf}. These
routines remove the contribution from astrophysical background, soft
proton background, and solar wind charge-exchange reactions. The final
net exposure times of the MOS1, MOS2, and pn cameras are 89.1, 92.5,
and 64.6~ks, respectively.

We created exposure-map-corrected, background-subtracted EPIC images
in three different energy bands: 0.3--1.1 (soft), 1.1--2.0 (medium),
and 2.0--8.0~keV (hard). Figure~\ref{fig:ngc2359_xrays} shows the
resulting images. These images have been adaptively smoothed using the
{\it adapt} task requesting 100, 50, and 50 counts for the soft,
medium, and hard band, respectively. As shown by Z14, diffuse emission
is detected towards the WR nebula, but we note here an additional
spatial component to the northeast of the central bubble. This is
spatially coincident with a blowout observed in
Figure~\ref{fig:image_ngc2359} and imaged in great detail as presented
in the Astronomy Picture of the Day by the team of the Star Shadows
Remote Observatory (SSRO) South on 2010 June 5\footnote{See
  http://apod.nasa.gov/apod/ap100605.html}. We will define the diffuse
X-ray emission from NGC\,2359 as the emission delineated by the solid
line aperture shown in Figure~\ref{fig:ngc2359_xrays}, which
encompasses the main nebula seen in Figure~1 and the extra blowout
detected towards the northeast.


Figure~\ref{fig:ngc2359_xrays} exhibits a significant number of
point-like sources projected within the WR nebula. We identify 16
sources within the diffuse emission, including the central WR star,
consistent with the pipeline identification of point-like sources. We
note that Z14 only excised the five point sources, marked by red and
green dashed circles in Figure~2 (bottom-left panel) from its spectral
analysis of the diffuse emission from NGC\,2359. Consequently, the
remaining 11 point sources contaminated the spectrum of the diffuse
emission presented by Z14.

\begin{figure}
\begin{center}
\includegraphics[width=1\linewidth]{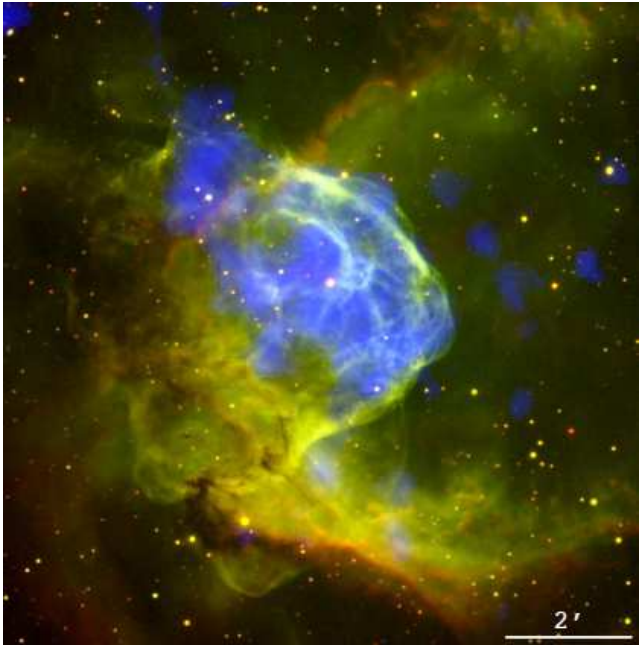}
\end{center}
\caption{Composite colour picture of the {\it XMM-Newton} observation
  of NGC\,2359. Red and green colours correspond to the H$\alpha$ and
  [O\,{\sc iii}] bands as shown in Figure~1 and the blue colour
  corresponds to the 0.3-1.1~keV soft emission. Point-like sources
  have been excised from this image.}
\label{fig:ngc2359_combined}
\end{figure}

In order to illustrate further the distribution of hot gas in
NGC\,2359 we have created a composite colour picture of the optical
and soft X-ray images as shown in
Figure~\ref{fig:ngc2359_combined}. Point-like sources have been
excised from this image making use of the CIAO {\it dmfilth} routine
\citep[CIAO Version 4.4;][]{Fruscione2006}. This image shows in great
detail that the hot gas in NGC\,2359 is distributed inside the main
bubble with a contribution filling the northeast blowout.

\section{Spectral extraction and analysis}
\label{sec:spectral_a}

For the rest of the X-ray data analysis, i.e., the spectra extraction
and modeling, we reprocessed the data with the SAS tasks {\it epproc}
and {\it emproc}. We then created lightcurves binning the data over
100~s for each of the EPIC cameras (MOS and pn) in the 10--12~keV
energy range to identify periods of high background level. The
background in the EPIC-pn and MOS cameras were considered high for
values of 0.5~counts~s$^{-1}$ and 0.2~counts~s$^{-1}$,
respectively. The net exposure times after excising high-background
levels are 98.6, 98.6 and 83.8~ks, for the MOS1, MOS2, and pn cameras,
respectively. As can be seen, this procedure does not reduce
considerably the net exposure time of the EPIC cameras as compared to
the net times obtained with the ESAS task, specially for the EPIC-pn
camera.

\begin{figure}
\begin{center}
\includegraphics[width=1\linewidth]{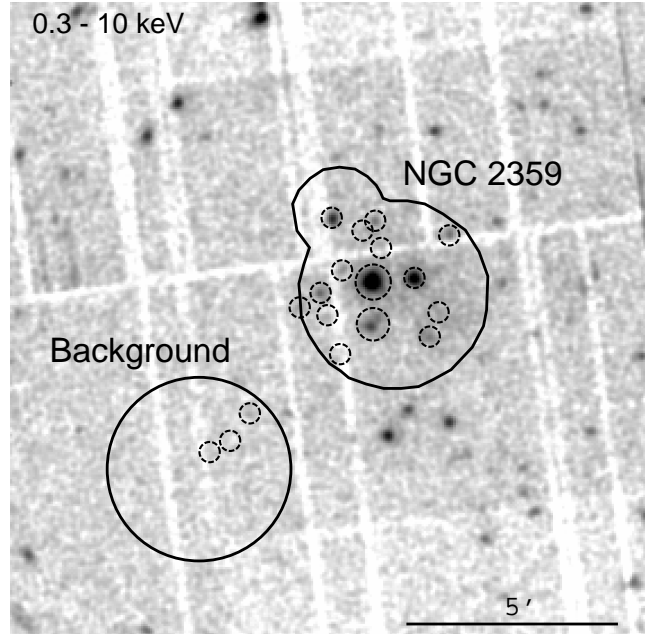}
\end{center}
\caption{EPIC (MOS1, MOS2, and pn) X-ray image of the field of view of
  NGC\,2359 in the energy range of 0.3-10~keV. The polygonal and
  circular apertures show the spectra subtraction regions (NGC\,2359
  and background). Dashed-line circular apertures show the position of
  the excised point-like sources.}
\label{fig:ngc2359_regions}
\end{figure}

The spectra of the three EPIC cameras and background have been
extracted from a polygonal and circular apertures, respectively, as
shown in Figure~\ref{fig:ngc2359_regions}. We note that the selection
of the background aperture by Z14 was unfortunate as it included
diffuse emission from the northeast blowout.

The count rates in the 0.3--1.5~keV range are 5.7, 6.1, and
25.9~counts~ks$^{-1}$, corresponding to 570, 600, and 2170 total
counts for the EPIC cameras MOS1, MOS2, and pn, respectively. In the
0.3--3.0~keV energy range, the corresponding count rates are 8.4, 8.8,
and 32.4~counts~ks$^{-1}$ and total counts 820, 870, and
2700~counts\footnote{Z14 only concentrated the spectral analysis in
  the 0.3-1.44~keV energy range.}. Figure~\ref{fig:pn_spectra}-left
shows the resultant background-subtracted spectra of the diffuse
emission in NGC\,2359 in the 0.3--3.0~keV energy range as extracted
from the EPIC-pn and MOS observations. A minimum of 60 counts per bin
was requested for the creation of the spectra.

To assess the contamination level to the diffuse emission caused by the 
11 point sources not excised by Z14, we have extracted their spectra from 
the EPIC-pn observations.
The total count number from these 11 point sources in the 0.3--3.0~keV 
energy range is $\approx$400~counts, i.e., about 15\% the count number 
of the diffuse X-ray emission in NGC\,2359. 
Remarkably, the average spectrum of these sources is much harder than that 
of the diffuse emission from NGC\,2359 (Figure~\ref{fig:ngc2359_xrays}), 
implying that their relative contribution to the spectrum extracted by Z14 
increases with energy.  
This modifies the spectral shape of the diffuse emission reported in 
that paper.

\subsection{Spectral properties}

\begin{figure*}
\begin{center}
\includegraphics[width=1.0\linewidth]{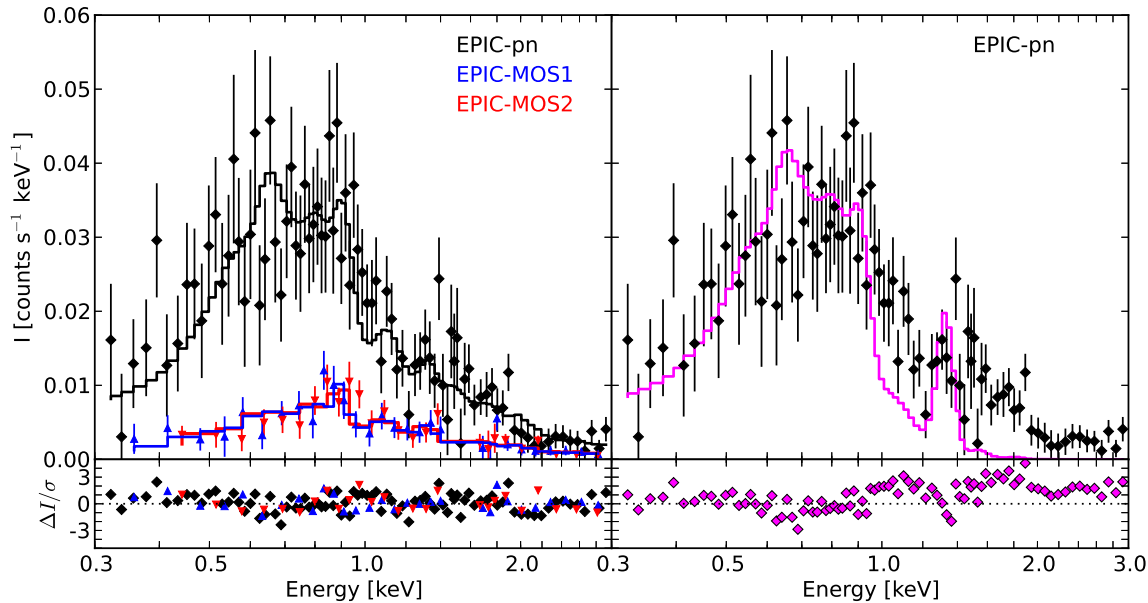}
\end{center}
\caption{Left: EPIC (pn, MOS1, and MOS2) spectra of NGC\,2359. The
  observed spectra are plotted in diamonds (black) and triangles (blue
  and red) for the pn and MOS cameras, respectively. Our best-fit
  model is plotted in solid black, blue, and red lines (Model E; see
  Table~1). Right: EPIC-pn spectrum (black diamonds) compared to the
  one-temperature plasma Model~B (magenta solid line).}
\label{fig:pn_spectra}
\end{figure*}

Figure~\ref{fig:pn_spectra} shows that the diffuse X-ray emission in
NGC\,2359 is detected mainly in the energy range of 0.3--2.0~keV, with
a broad peak around 0.5--0.9~keV. More specifically, there are two
apparent maxima at 0.6--0.7 and at 0.9~keV. There is a sharp decline
above 0.9~keV with line contribution at $\lesssim$1.5~keV which would
correspond to the Mg\,{\sc xi} line described by Z14.

Following previous X-ray studies of WR nebulae, we used XSPEC
\citep[v12.8.1;][]{Arnaud1996} to model the diffuse X-ray emission in
NGC\,2359. We first tried to fit an absorbed one-temperature {\it
  apec} optically thin plasma emission model following the procedure
described by Z14, adopting nebular abundances as those reported by
\citet{Esteban1993}. To allow a fair comparison we fixed the column
density to $N_\mathrm{H}$=4.77$\times$10$^{21}$~cm$^{-2}$, which is
the value adopted in Z14, even though this value is higher than the
upper limit indicated by the extinction towards the central star in
NGC\,2359, WR\,7. Details of all models attempted in this section are
presented in Table~1.

The first model was performed on the EPIC-pn spectrum (Model~A). This
resulted in a poor fit ($\chi^{2}$/DoF=3.30) with a temperature
component of $kT$=0.180~keV. The second model was attempted with
variable Mg (Model~B) as used by Z14 but resulted in a similarly poor
fit ($\chi^{2}$/DoF=2.96; $kT$=0.175~keV) and a magnesium abundance of
$\mathrm{X_{Mg}}$=4$\pm$0.6 times its solar value
\citep{Anders1989}. This model is presented in
Figure~\ref{fig:pn_spectra}-right panel with a magenta solid line in
comparison with the EPIC-pn spectrum (black diamonds).

We next model the EPIC-pn spectrum with two temperature components
(Model~C). This resulted in a better fit ($\chi^{2}$/DoF=1.01) with
plasma temperatures of $kT_{1}$=0.16~keV and $kT_{2}$=4.6~keV. Model~D
allowed the magnesium abundance to vary, but it resulted in very
similar temperature components and $\mathrm{X_{Mg}}$ varied around its
solar value ($\mathrm{X_{Mg}}$=1.75$^{+1.30}_{-1.20}$).

\begin{table*}
\footnotesize
\centering
\caption{Spectral fits of the diffuse X-ray emission in NGC\,2359}
\begin{tabular}[1.0\textwidth]{cccccccccr}
\hline
\hline
Model    & $kT_{1}$               &  $A_{1}^\mathrm{a}$      &  $f_{1}^\mathrm{b}$     & $F_{1}^\mathrm{b}$     &   $kT_{2}$        & $A_{2}^\mathrm{a}$      & $f_{2}^\mathrm{b}$      & $F_{2}^\mathrm{b}$ &$\chi^{2}$/DoF \\
         &  (keV)                 & (cm$^{-5}$)             &  (cgs)                &(cgs)                  & (keV)            &(cm$^{-5}$)              &(cgs)                  &(cgs)              & \\
\hline
  A      & 0.180$^{+0.002}_{-0.002}$&  2.26$\times$10$^{-3}$  & 4.2$\times$10$^{-14}$  & 1.2$\times$10$^{-12}$  &  \dots               & \dots       & \dots                       & \dots        &280.82/58.10=3.30\\
  B      & 0.175$^{+0.002}_{-0.002}$ & 2.20$\times$10$^{-3}$ & 4.6$\times$10$^{-14}$  & 1.2$\times$10$^{-12}$  &  \dots               & \dots       & \dots                       & \dots        &249.20/84.19=2.96\\
\hline
  C      & 0.161$^{+0.013}_{-0.010}$ &2.70$\times$10$^{-3}$  & 3.5$\times$10$^{-14}$   &1.2$\times$10$^{-12}$   &4.6$^{+3.2}_{-1.4}$ & 1.10$\times$10$^{-4}$ & 3.0$\times$10$^{-14}$ & 1.1$\times$10$^{-13}$ &84.13/83=1.01\\
  D      & 0.160$^{+0.013}_{-0.011}$ &2.72$\times$10$^{-3}$  & 3.6$\times$10$^{-14}$   &1.2$\times$10$^{-12}$   &4.5$^{+3.2}_{-1.5}$ & 1.05$\times$10$^{-4}$ & 2.9$\times$10$^{-14}$ & 8.0$\times$10$^{-14}$ &83.09/82=1.01\\
  E      & 0.173$^{+0.003}_{-0.018}$ &2.04$\times$10$^{-3}$  & 3.5$\times$10$^{-14}$   &1.1$\times$10$^{-12}$   &4.9$^{+2.4}_{-1.3}$ & 1.07$\times$10$^{-4}$ & 2.9$\times$10$^{-14}$ & 8.0$\times$10$^{-14}$ &125.92/135=0.93\\
\hline
  F$^\mathrm{c}$      & 0.157$^{+0.020}_{-0.011}$ &3.60$\times$10$^{-3}$  & 3.7$\times$10$^{-14}$   &1.6$\times$10$^{-12}$   &4.8$^{+2.2}_{-1.4}$ & 1.10$\times$10$^{-4}$ & 2.8$\times$10$^{-14}$ & 8.3$\times$10$^{-14}$ &125.47/134=0.94\\
\hline
\hline
\end{tabular}
\begin{list}{}{}
\item{ $^{\mathrm{a}}$The normalization parameters is defined as
    $A=1\times10^{-14}\int n_\mathrm{e}n_\mathrm{H}dV/4 \pi d^{2}$,
    where $d, n_\mathrm{e}$, and $V$ are the distance, electron
    density, and volume in cgs units, respectively.}
\item{ $^{\mathrm{b}}$$f$ and $F$ represent the absorbed and
    unabsorbed fluxes. Fluxes are computed in the 0.3--2.0~keV energy
    range. All fluxes are in cgs units (erg~cm$^{-2}$~s$^{-1}$).}
\item{$^{\mathrm{c}}$Model~F was performed with a variable absorption
    column density. This model resulted in a
    $N_\mathrm{H}=$5.15$\times$10$^{ 21}$~cm$^{-2}$.}
\end{list}
\end{table*}

We decided to fit simultaneously the three EPIC (pn, MOS1, and MOS2)
spectra\footnote{Note that the count number in our MOS spectra are
  comparable to that find for the EPIC-pn spectrum in Z14.} with a
two-temperature optically thin plasma model and abundances as those
determined from the optical nebula (Model~E). The resultant plasma
temperature components are $kT_{1}$=0.17~keV and $kT_{2}$=4.9~keV with
a good quality ($\chi^{2}$=0.93). Model~E is shown in
Figure~\ref{fig:pn_spectra}-left panel with solid black, blue, and red
lines for the EPIC-pn, MOS1, and MOS2, respectively.

Finally, we fit the three EPIC cameras with a two-temperature plasma
model with variable absorption column density (Model~F). This also
resulted in a good fit ($\chi^{2}$/DoF=0.94) with a higher absorption
column, $N_\mathrm{H}=(5.15^{+0.7}_{-0.4})\times$10$^{ 21}$~cm$^{-2}$,
and similar temperature components within the errors
($kT_{1}$=0.157~keV and $kT_{2}$=4.75~keV; see Table~1) as Model~E.

We would like to mention that models with variable neon abundance were
also attempted to raise the emission around 0.9~keV, but did not
yield a major improvement, with values ranging around its nebular
value.

\section{Discussion}

We have reanalysed the {\it XMM-Newton} archival observations towards
NGC\,2359 in two steps:
(i) using the XMM-ESAS task to study the distribution of the diffuse
X-ray emission and point-like sources and (ii) extracting spectra from
the three EPIC cameras reprocessing the observations with the classic
SAS procedures.

The X-ray images of NGC\,2359 allowed us to disentangle the diffuse
emission from that coming from point-like sources in a very effective
way. This procedure has also helped us identify an additional spatial
component associated with a blowout at the northeast rim of the WR
bubble. Such blisters or blowouts are also detected in the {\it
  XMM-Newton} and {\it Chandra} observations towards S\,308 and
NGC\,6888 \citep[][and Toal\'{a} et al. in prep.]{Toala2012,Toala2014}
and might have different origins. In the case of NGC\,6888 the caps
seem to be interacting with the interstellar medium and the formation
of the blowout is due to the low density towards the northwest
\citep[see Figure~7 in][]{Toala2014}. In the case of NGC\,2359, it
seems to be interacting with material ejected in a previously eruptive
and non-isotropic wind, this is, the previous dense material did not
have a $\rho \sim r^{-2}$ distribution.  Indeed, \citet{Rizzo2003}
found three different velocity components towards NGC\,2359 that
reinforces the idea that WR\,7 evolved through an LBV
phase. Furthermore, it is probable that the southeastern blister shown
in Figure~\ref{fig:image_ngc2359} is also part of the main nebula but
due to the molecular material in the line of sight towards this region
\citep[][]{StLouis1998,Cappa2001,Rizzo2001,Rizzo2003} could not be
detected in X-rays. This is corroborated by the lack of diffuse X-ray
emission towards the southeast of the nebula as illustrated in
Figure~\ref{fig:ngc2359_xrays} and \ref{fig:ngc2359_combined}. The
variations in the foreground absorption precludes a clear view of the
distribution of the X-ray-emitting gas. With all this in mind, the
formation scenario of such a complex WR bubble should not be taken
lightly as it would require the assumption of a previous LBV
non-isotropic wind with massive ejections of material and the
inclusion of the photoionizing flux from the central star in the WR
phase.

The physical conditions of the hot gas in NGC\,2359 can be assessed by
modeling the X-ray spectra. Table~1 and Figure~\ref{fig:pn_spectra}
illustrate that one-temperature plasma models do not result in good
quality fits. A one-temperature plasma model is not able to fit the
most energetic part of the spectrum as shown by the magenta line in
Figure~\ref{fig:pn_spectra}-right panel (Model~B). The analysis
presented by Z14 does not model the spectral range above 1.44~keV, in
which a considerably emission is present. Leaving the magnesium
abundance as a free parameters makes the abundance increase because
the model tries to compensate the lack of emission in the
one-temperature model towards energies around $\lesssim$1.5~keV. On
the other hand, two-temperature plasma models improve the fits to the
observed spectra and it is consistent to that found for other WR
nebulae
\citep[e.g.,][]{Chu2003,Zhekov2011,Toala2012,Toala2013,Toala2014}. A
low temperature component is used to model the bulk of the X-ray
emission while a higher-temperature component models the extra
emission at higher energies. Model~E, performed with a fixed
$N_\mathrm{H}=$4.77$\times$10$^{ 21}$~cm$^{-2}$ and simultaneously
fitting the three EPIC spectra, yielded reasonable fits with
temperature components $T_{1}$=2$\times$10$^{6}$~K and
$T_{2}$=5.7$\times$10$^{7}$~K and $X_\mathrm{Mg}$ close to the solar
value. Model~F, which allowed the absorption column to vary, resulted
in plasma components of $T_{1}$=1.8$\times$10$^{6}$~K and
$T_{2}$=5.6$\times$10$^{7}$~K with $N_\mathrm{H}=$5.15$\times$10$^{
  21}$~cm$^{-2}$. This last model was attempted because the absence of
diffuse X-ray emission towards the southeast of the main bubble could
be due to higher absorption column densities.

Models~E and F appear to indicate that the plasma in NGC\,2359 has
higher temperatures than those reported for S\,308 and NGC\,6888, but
in accordance with those WR bubbles, the flux ratio ($F_{1}/F_{2}$)
show that the secondary component represents $<$10\% of the observed
emission in NGC\,2359.
The total observed flux in the 0.3--2.0~keV energy range of Model~E is
$f_\mathrm{X,E}$=6.4$\times$10$^{-14}$~erg~cm$^{-2}$~s$^{-1}$ which
corresponds to an unabsorbed flux of
$F_\mathrm{X,E}$=1.1$\times$10$^{-12}$~erg~cm$^{-2}$~s$^{-1}$. The
estimated observed flux from Model~F is the same but due to its higher
absorption column density, its unabsorbed flux is
$F_\mathrm{X,F}$=1.3$\times$10$^{-12}$~erg~cm$^{-2}$~s$^{-1}$. The
estimated luminosities for these flux values at a distance of 3.67~kpc
\citep{vanderH2001} are
$L_\mathrm{X,E}$=1.8$\times$10$^{33}$~erg~~s$^{-1}$ and
$L_\mathrm{X,F}$=2.0$\times$10$^{33}$~erg~~s$^{-1}$. Finally, using
the normalization parameters of best-fit models we can estimate the
electron density of the X-ray-emitting gas in NGC\,2359 assuming a
spherical morphology with a radius 2.2$'$. This sets an upper value to
the electron density as $n_\mathrm{e}\lesssim0.6$~cm$^{-3}$.

As in other wind-blown bubbles (e.g., planetary nebulae and H\,{\sc
  ii} regions), the hot bubble in NGC\,2359 can be expected to have a
temperature $>$10$^{7}$~K as calculated for an adiabatically shocked
gas \citep[e.g.,][]{Dyson1997} for a stellar wind velocity of
1600~km~s$^{-1}$ \citep{vanderH2001}. The gas inside the hot bubble is
very tenuous ($n_\mathrm{e}$=10$^{-3}$--10$^{-2}$~cm$^{-3}$), thus it
would produce low-luminosity, hard X-rays non detectable by the
current X-ray satellites. \citet{Weaver1977} proposed that for the
case of interstellar bubbles, thermal conduction at the interface
between the hot bubble and the outer ionized nebula could diminish the
temperature of the hot bubble while raising its density. This model,
however, fails in explaining the low X-ray luminosities reported by
observations of diffuse X-ray-emitting gas in planetary nebulae, WR
bubbles, and superbubbles around young star clusters \citep[e.g.,][and
references therein]{Townsley2003,Toala2012,Ruiz2013}. Thus, none of
these two models are applicable to the diffuse X-ray emission in
wind-blown bubbles.  The fragmentation of the ionized shell due to
instabilities and the effects of the ionization flux from the central
star needs to be incorporated in the models.  The clumps formed by
instabilities in the wind-wind interaction are an important source of
mass, injecting material into the hot bubble via hydrodynamic ablation
and photoevaporation. Two-dimension numerical simulations taking into
account the time evolution of the stellar wind parameters and
ionization photon flux are able to give a reasonable description of
both the plasma temperatures and luminosities in WR nebulae
\citep[][]{Toala2011,Dw2013}.  Thus, the estimated plasma temperature,
electron density, luminosity, and nebular abundances of the diffuse
X-ray emission in NGC\,2359 indicate that the density in the hot
bubble has been raised by mixing of material from the outer nebula as
in S\,308 and NGC\,6888.

\section{Conclusions}

We have presented our analysis of archival {\it XMM-Newton}
observations of NGC\,2359. We find significant differences between our
analysis and that presented in Z14. Our findings can be summarized as:

\begin{itemize}

\item There is an additional spatial component to the diffuse X-ray
  emission towards the northeast from the main bubble in NGC\,2359
  identified as a blister in optical images.\\

\item We identify 16 point-like sources projected in the line of sight
  of the WR nebula, including WR\,7. We have excised all of these
  sources for a cleaner spectral analysis.\\

\item Our background selection in the spectral analysis does not
  contain any contribution of the diffuse X-ray emission from a
  blowout nor point-like sources, allowing us to perform a more
  accurate spectral analysis in the 0.3-2.0~keV energy range.\\

\item We model the diffuse X-ray emission with a two-temperature
  plasma model for abundances of the optical WR nebula without
  magnesium enhancement. This model resulted in plasma temperatures of
  $T_{1}$=2$\times$10$^{6}$~K and $T_{2}$=5.7$\times$10$^{7}$~K. The
  second temperature component contributes less than 10\% of the total
  flux in the 0.3--2.0~keV energy range, similar to other WR
  bubbles.\\

\item The estimated unabsorbed flux and X-ray luminosity of NGC\,2359
  are $F_\mathrm{X}$=1.3$\times$10$^{-12}$~erg~cm$^{-2}$~s$^{-1}$ and
  $L_\mathrm{X}$=2$\times$10$^{33}$~erg~~s$^{-1}$, respectively.\\

\item The estimated electron density,
  $n_\mathrm{e}\lesssim$0.6~cm$^{-3}$, and the fact that the X-ray
  emission can be modeled with the abundances as those as the optical
  nebula imply that the hot bubble in NGC\,2359 has raised its density
  as a result of strong mixing from the outer material.

\end{itemize}

\section*{Acknowledgments}

The authors would like to thank Steve Mazlin, Jack Harvey, Daniel
Verschatse, and Rick Gilbert from SSRO-South and PROMPT/CTIO for
providing the optical images of NGC\,2359. J.A.T. acknowledges support
by the CSIC JAE-Pre student grant 2011-00189. J.A.T. and M.A.G. are
supported by the Spanish MICINN grant AYA 2011-29754-C03-02 co-funded
with FEDER funds.

\end{document}